\documentclass{jpp}

\usepackage{graphicx}
\usepackage{natbib}
\usepackage[colorlinks,urlcolor=blue,citecolor=blue,linkcolor=blue]{hyperref}
\usepackage{amsmath}
\usepackage{subcaption}
\usepackage{gensymb}

\ifCUPmtlplainloaded \else
  \checkfont{eurm10}
  \iffontfound
    \IfFileExists{upmath.sty}
      {\typeout{^^JFound AMS Euler Roman fonts on the system,
                   using the 'upmath' package.^^J}%
       \usepackage{upmath}}
      {\typeout{^^JFound AMS Euler Roman fonts on the system, but you
                   dont seem to have the}%
       \typeout{'upmath' package installed. JPP.cls can take advantage
                 of these fonts, if you use 'upmath' package.^^J}%
      }
  \else
  \fi
\fi

\ifCUPmtlplainloaded \else
  \checkfont{msam10}
  \iffontfound
    \IfFileExists{amssymb.sty}
      {\typeout{^^JFound AMS Symbol fonts on the system, using the
                'amssymb' package.^^J}%
       \usepackage{amssymb}%
       \let\le=\leqslant  \let\leq=\leqslant
       \let\ge=\geqslant  \let\geq=\geqslant
      }{}
  \fi
\fi

\ifCUPmtlplainloaded \else
  \IfFileExists{amsbsy.sty}
    {\typeout{^^JFound the 'amsbsy' package on the system, using it.^^J}%
     \usepackage{amsbsy}}
    {\providecommand\boldsymbol[1]{\mbox{\boldmath $##1$}}}
\fi

\usepackage{graphicx}
\usepackage{wrapfig}
\usepackage{natbib}
\usepackage{url}
\usepackage{enumitem}

\usepackage{graphics}

\newcommand{\be}{\begin{equation}}
\newcommand{\ee}{\end{equation}}

\newcommand{\ba}{\begin{eqnarray}}
\newcommand{\ea}{\end{eqnarray}}
\newcommand{\om}{\omega}
\newcommand{\Alfven}{Alfv\'{e}n }

\newcommand{\LC}{light cylinder}
\newcommand{\Bf}{{magnetic field}}

\newcommand{\NS}{neutron star}
\newcommand{\NSs}{{neutron stars}}
\newcommand{\EM}{electromagnetic}

\newcommand{\ms}{magnetosphere}
\newcommand{\mss}{magnetospheres}

\newcommand{\Lf}{Lorentz factor}

\title{Free electron laser in the magnetically dominated regime: simulations with the ONEDFEL code}
\author{Henry Freund$^{(1)}$, Maxim Lyutikov$^{(2)}$}

\author{ Maxim Lyutikov \aff{1}
  \and Henry Freund  \aff{2}}
\affiliation{\aff{1} 
Department of Physics and Astronomy, Purdue University, 525 Northwestern Avenue, West Lafayette, IN, 47907-2036, USA \\
\aff{2} Institute for Research in Electronics and Applied Physics, University of Maryland, College Park, MD 20742
and 
Department of Electrical and Computer Engineering, University of New Mexico, Albuquerque, NM 87131
}

\begin{document}

\maketitle

\begin{abstract}
Using  the ONEDFEL code we perform  Free Electron Laser simulations in the  astrophysically important   guide-field dominated regime. 
For wigglers' (\Alfven waves)  wavelengths of  tens of kilometers  and beam \Lf\ $\sim 10^3$, the  resulting  coherently emitted  waves  are in the centimeter range.
Our simulations show  a growth of  the wave intensity over fourteen orders of magnitude,  over the  astrophysically relevant scale of $\sim$ few kilometers. The signal grows from noise (unseeded). The resulting spectrum shows fine spectral sub-structures, reminiscent  of the ones observed in Fast Radio Bursts (FRBs).
\end{abstract}


\section{Introduction: Free Electron Laser in  astrophysical setting}

Pulsars' radio  emission mechanism(s)   eluded identification for nearly half a century \citep[\eg][]{Melrose00Review,2008AIPC..983...29L,2016JPlPh..82c6302E}.  Most likely, several types of coherent processes operate in different  sources (\eg magnetars versus pulsars), and in different parts of pulsar \mss\ \citep[see \eg\ discussion in][]{2016MNRAS.462..941L}.  

The problem of pulsar coherent emission generation  has been brought back to the research  forefront by the meteoritic  developments over the last years  in the field of mysterious Fast Radio Bursts (FRBs). Especially important was the detection of a radio burst from a Galactic magnetar by CHIME and STARE2 collaborations in coincidence with high energy bursts \citep{2020Natur.587...54C,2021NatAs...5..372R,2020Natur.587...59B,2020ApJ...898L..29M,2021NatAs.tmp...54L}.  The similarity  of  properties of magnetars' bursts   to the Fast Radio Bursts gives credence to the magnetar origin of FRBs (even though the radio  powers are quite different - there is a broad distribution).

The phenomenon of Fast Radio Bursts challenges our understating of relativistic plasma coherent processes to the extreme. In this case radio waves can indeed carry an astrophysically important amount of the energy. For example, radio luminosity in FRBs can match, for a short period of time, the macroscopic Eddington luminosity and exceed total Solar luminosity by many orders of magnitude. Still, the fraction emitted in radio remains small -  this   relatively  small  fraction of total energy  that pulsars and FRBs emit in radio ($\sim 10^{-5}$ is typical) is  theoretically challenging: simple order-of-magnitude estimates cannot be used. Emission production and saturation levels of instabilities  depend on the kinetic details of the plasma  distribution function.

\cite{2021ApJ...922..166L}   developed a model of the generation of coherent radio emission in the Crab pulsar, magnetars and Fast Radio Bursts (FRBs) due to   a variant of the free-electron laser (FEL) mechanism, operating in a weakly-turbulent, guide-field dominated plasma.
This  presents a new previously unexplored way (in astrophysical settings) of producing coherent emission  via {\it parametric} instability.

A particular regime of the FEL (SASE - Self-Amplified Spontaneous Emission) 
micro-bunching is initiated by the spontaneous radiation. 
In the beam frame the wiggler and the \EM\ wave have the  same frequency/wave number, but propagate in the opposite direction. The addition of two counter-propagating waves creates a standing wave in the  beam frame. The radiation energy density is smaller at the nodes of the standing  wave: this creates a ponderomotive force that pushes the particles towards the nodes -  bunches are created.  These bunches are still shaken by the \EM\ wiggler: they emit in phase, coherently.

Somewhat surprisingly,  the  FEL model  in magnetically dominated regimes  \citep{2021ApJ...922..166L} is both robust to the underlying plasma parameters and succeeded in reproducing a number of subtle observed features: (i) emission frequencies depend mostly on the scale    of turbulent fluctuations and the Lorentz factor of the reconnection generated beam,  Eq. (\ref{omC});  it is independent of the absolute value of the underlying magnetic field. (ii) The model explained both broadband emission and the presence of emission stripes, including multiple stripes observed in the High Frequency Interpulse of the Crab pulsar. (iii) The model reproduced correlated spectrum-polarization properties: the  presence of narrow emission bands in the spectrum favors linear polarization, while broadband emission can have arbitrary polarization. The model is applicable to a 
very broad range of \NS\ parameters: the model is mostly  {\it independent} of the value of the \Bf. It is thus applicable to a broad variety of NSs, from  fast spin/weak \Bf\ millisecond pulsars to slow spin/super-critical   \Bf\ in magnetars, and from regions near the surface up to (and a bit beyond of) the \LC.

The   guide field dominance plays a tricky role in the operation of an  FEL. On the one hand it suppresses the growth rate. But what turns out to be more important in astrophysical applications 
is that  the guide field dominance 
  helps to maintain beam coherence.
   Without the guide field,  particles with different energies follow different trajectories in the \Bf\ of the wiggler, and quickly lose coherence even for small initial velocity spread. In contrast, in the guide-field dominated regime all particles follow, basically, the same trajectory. Hence coherence is maintained as long as the velocity spread {\it in the beam frame} is $\Delta \beta \leq 1$.  Such tolerance to velocity spread is an unusual property of guide-field dominaed FEL.

\section{FEL in guide-field dominated regime: theoretical summary}

\subsection{Model parameters}

Let us next discuss the basic model parameters. (Unfortunately, there is some confusion in standard definitions used in literature.)

The model starts with an assumption that guiding  \Bf\ lines are perturbed by a packet of linearly polarized   \Alfven waves of intensity $B_w$ and frequency $\om$.
 In highly magnetized  force-free plasma \Alfven waves propagating along the \Bf\ are nearly luminal, $v_A \approx c$. 

The first parameter is  dimensionless wave intensity. We chose notation  $a_0$,  which is standard in the laser community and sometimes used in the FEL community as well,
\be
a_0 = \frac{e B_w}{m_e c \om} \gg 1
\ee
It should be remarked that in the FEL literature parameters $K$ and $a_0$ are often used interchangeably.  In this paper, we follow the nomenclature used for $K$ in Eq. (2.4) below in \cite{1975clel.book.....J}, parag. 14.7

For guide-field dominated regime another parameter is  the  relative  intensity
\be
\delta= \frac{B_w}{B_0}
\ll 1
\ee
where $B_0$ is a guide field. We assume that the  wave is relatively weak, $\delta \ll 1$.

  A (reconnection-generated) beam of particles with \Lf\ $\gamma_b$ propagates along the rippled \Bf\ in a direction opposite to the direction of \Alfven  waves. In the frame of the beam the waves are seen with $k_{w}'= 2 \gamma_b k_{w} $. In the guide-field dominated regime the cyclotron frequency  associated with the guide field is much larger than the 
   frequency of the wave in the beam frame,  and the  cyclotron frequency  associated with the fluctuating field; hence
  \be
\Omega_0 \gg   k_{w}' c , \Omega_w 
  \ee
where $\Omega_0 = e B_0/(m_e c) $ is the cyclotron frequency (non-relativistic) of the guide field, $\Omega_w = e B_w/(m_e c) $ is the cyclotron frequency associated with the wiggler field.  

  Another important parameter,  defined by    \cite[][parag. 14.7] {1975clel.book.....J}   is wiggler-undulator parameter
    \be
  K = \delta \gamma_b
  \label{K} 
  \ee
  This parameter is related to the magnitude of the wiggler-induced oscillations in the beam trajectory, which is also related to the opening angle of the cone of the generated radiation. When $ K \gg  1$  this oscillation is large and the pump field is sometimes referred to as a ``wiggler''. In the opposite regime where $ K \ll  1$  the pump field sometimes referred to as an ``undulator''.  In this paper, we will refer to the pump field as a wiggler throughout. 
  
\begin{figure}
  \centering
    \includegraphics[width=0.49\textwidth]{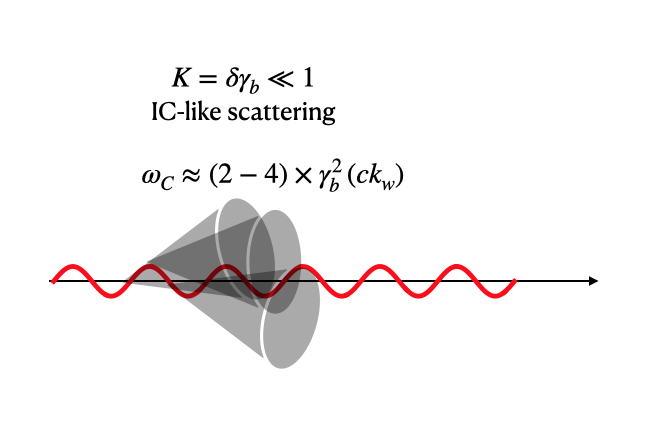}
      \includegraphics[width=0.49\textwidth]{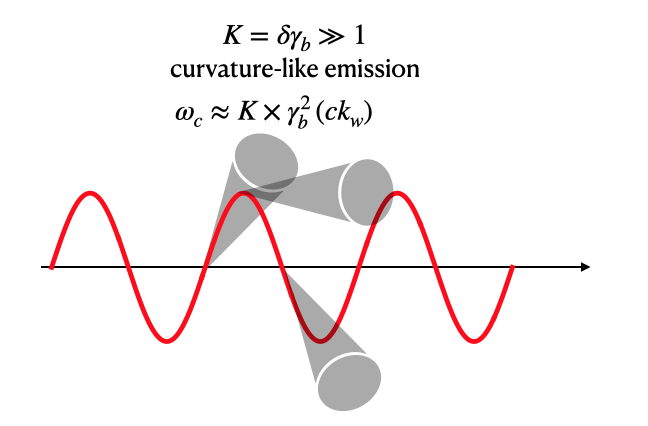}
     \caption{\bf Different regimes of wave-particle interaction depending on the parameter  $K$ (\ref{K}). Gray cones illustrate instantaneous  emisison cone.}
\label{IC-like}
\end{figure}

   
   The $K$ parameter (\ref{K})  is a product of two quantities, relative amplitude $\delta \ll 1$ and \Lf\ $\gamma_ b \gg 1$, so generally its values can be either  large and small.    A relativistically moving electron  emits in a cone with opening angle $\Delta \theta \sim 1/\gamma$. 
In the 
 $K \ll 1$ regime  regime that opening angle is much larger than the variation in the bulk direction of emission at different points in the trajectory,  Fig. \ref{IC-like}, left panel.  The radiation detected by an observer is an almost coherent superposition of the contributions from all the oscillations in the trajectory at a frequency
 \be
\om_C = 
 \gamma_b^2 ( c k_{w}) \times
  \left\{ \begin{array}{cc}
4, & \mbox{EM wiggler}
\\
2,& \mbox{static wiggler}
\end{array}  \right.
\label{omC} 
\ee
We note that the axial velocity in the strong axial guide field regime is nearly constant and close to the speed of light for high energy electrons, see Fig. \ref{001} in \S \ref{CARM}. Hence, the resonant frequency is also nearly constant.
 
In the $K\gg 1$ regime the variations in the direction of emission is much larger than the angular width of the emission at any point in the trajectory,  Fig. \ref{IC-like}, right panel.  As a result, an observer located within angle $\leq \delta$ with respect to the overall guide field see periodic bursts of emission with typical frequency
\be
\om_c \approx  \gamma_b^2 K  \times  ( c k_{w})
\ee
Since $ K \gg  1$ in this regime, the resulting   frequency is higher than that for the case where $ K \ll 1$. In this paper, we work in the $ K \ll 1$ regime which is the regime more typical of FELs.

 Another terminology issue: here we contrast  the term ``Compton''  with ``curvature'', not with ``Raman'' regime, which is Compton-like scattering, but on collective plasma fluctuations.

In this paper we work in the regime $K \ll 1$ (this is the usual regime of FELs). The scattered frequency is then given by (\ref{omC}); below  we drop the subscript $C$.

\subsection{Overview of main results of \cite{2021ApJ...922..166L}}

\cite{2021ApJ...922..166L}  developed a model of the generation of coherent radio emission in the Crab pulsar, magnetars and Fast Radio Bursts (FRBs) whereby the  emission is produced by a reconnection-generated beam of particles via a variant of Free Electron Laser (FEL) mechanism, operating in a weakly-turbulent, guide-field dominated plasma. The guide-field dominanbce is a key new feature that distiguishes this regime from the conventional laboratory FELs:   a (reconnection-generated) beam of particles with \Lf\ $\gamma_b$ propagates along the wiggled \Bf\ in a direction opposite to the direction of \Alfven  waves. In the frame of the beam the waves are seen with $k_{w,b}= 2 \gamma_b k_{w} $. Guide field dominance requires $2 \gamma_b k_{w} c \ll \Omega_0$.

The key  results  of   \cite{2021ApJ...922..166L} are as follows:
\footnote{ Notations used in   \cite{2021ApJ...922..166L}  are somewhat different: $a_H \to \delta$, $a_A \to a_0$, $a_{H,b} \to K= \gamma_b \delta$, $\om_B \to \Omega_0$,
 $\delta \to  a_0 \delta  \gamma_b \Omega_0$.}
\begin{itemize} 
\item The  interaction Hamiltonian (this is    particularly  important  for the present work, as it  describes evolution of the instability and its saturation).
Particle motion in the combined fields of the wiggler $B_w$, the  EM wave $E_{EM}$ (both with wave vector $k_{w}' $ in the beam frame) and the    guide-field  $B_0 \gg E_w, E_{EM}$ can be described by a simple  ponderomotive Hamiltonian \cite[Eq. (39)][]{2021ApJ...922..166L}
\be
{\cal H} = \frac{\beta_z^2}{2} +  \delta \left(  \frac{ E_{EM}}{B_0}  \right)\left(  \frac{\Omega_0}{ k_w c}  \right) \sin ^2(  k_{w}'  z)
\label{H}
\ee
where $\beta _z $ is axial velocity, see Fig. \ref{X-Y-Z-3D}.  The second term is the ponderomotive potential. The  Hamiltonian formulation allows powerful analytical methods to be applied to the system (adiabatic invariant, phase space separatrix etc).  This is especially important for the estimates of the non-linear saturation, one of the main goals of the present work.
\begin{figure}
\centering
\includegraphics[width=.49\textwidth]{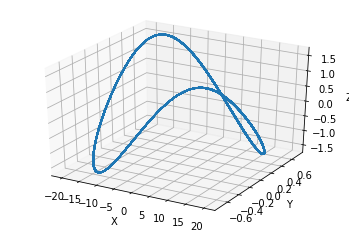} 
\includegraphics[width=.49\textwidth]{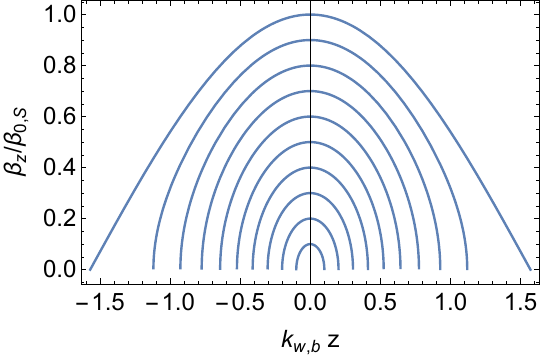} 
\caption{ Particle dynamics for FEL in the guide-field dominated plasma with the ponderomotive potential (\ref{H}) \protect\cite[arbitrary unites chosen to illustrate the trajectory][]{2021ApJ...922..166L}. Left panel: 3D rendering of particles trajectories in the beam frame (a particle experiences a saddle-like trajectory). Right  panel:  trajectory of trapped particles in phase  $\beta_z-z$ plane; velocity is normalized to the separatrix $\beta_{S} $, Eq. (\ref{vbetaS}). }
\label{X-Y-Z-3D} 
\end{figure}

\item The  growth rate  of the parametric instability  is  \cite[][Eq. (62)]{2021ApJ...922..166L} 
\be
\Gamma = \left(   \left(   \frac{ E_w}{B_0}    \right) \left(  \frac{ E_{EM}}{B_0}  \right) {\Omega_0}{ k_w c}  \right)^{1/2} \propto B_0^{-1/2}
\label{Gamma}
\ee
Importantly, it  is  only mildly suppressed by the strong guide field.
\item Saturation  level.  The ponderomotive potential increases linearly with EM wave intensity,  while energy density of EM waves increases quadratically. The corresponding  saturated velocity jitter   \cite[][Eq. (73)]{2021ApJ...922..166L} 
\be
\beta_{S} =  \frac{\delta}{ \sqrt{2 \gamma}} \frac{\om_p}{k_w c}
\label{vbetaS} 
\ee
\item The guide field dominance plays a dual role. First, it reduced the growth rate, but only mildly, $\propto B_0^{-1/2}$, Eq. (\ref{Gamma}). What is more important, the strong guide field  helps to maintain beam coherence. Without the guide field,  particles with different energies follow different trajectories, and quickly lose coherence even for small initial velocity spread. In contrast, in the guide-field dominated regime all particles follow, basically, the same trajectory. Hence coherence is maintained as long as the velocity spread  in the beam frame is $\Delta \beta \leq 1$ \citep[Fig. 13 of][]{2021ApJ...922..166L}. As a result, the model require only a  mildly narrow distribution of the beam's particles, $\Delta p /p_0 \leq 1$ and the spectrum of turbulence  $\Delta k_{w,b}/k_{w,b} \leq 1$
\item  The model operates in a very broad range of \NS's parameters: the model is independent of the value of the \Bf. It is thus applicable to a broad variety of NSs, from  fast spin/weak \Bf\ millisecond pulsars to slow spin/super-critical   \Bf\ in magnetars, and from regions near the surface up to (and a bit beyond of) the \LC.
\end{itemize}

\section{Simulations with ONEDFEL codes}
\label{MINERVA}

\subsection{The codes}

In this work we  performed simulations of the 
  interaction of a single-charged relativistic beam with the wiggler using FEL code ONEDFEL  \citep{freund1986principles}.
  ONEDFEL is time-dependent code that simulates the FEL interaction in one-dimension. The radiation fields are tracked by integration of the wave equation under the slowly-varying envelope approximation. As such, the wave equation is averaged over the fast time scale  under the assumption that the wave amplitudes vary slowly over a wave period. The dynamical equations are a system of ordinary differential equations for the mode amplitudes of the field and the Lorentz force equations for the electrons which are integrated simultaneously using a 4th order Runge-Kutta algorithm. Time dependence is treated by including multiple temporal ``slices'' in the simulation which are separated by an integer number of wavelengths. The numerical procedure is that each slice is advanced from  $z \to  z + \Delta z$ separately by means of the Runge-Kutta algorithm. Time dependence is imposed as an additional operation by using the forward time derivative as an additional source term to treat the slippage of the radiation field with respect to the electrons.
 Slippage occurs at the rate of one wavelength per undulator period in the low gain regime and after saturation of the high gain regime but at the rate of  one third of a wavelength per undulator period in the exponential gain regime \citep{1990NCimR..13i...1B,1995PhR...260..187S,freund1986principles}. ONEDFEL self-consistently describes slippage in each of these regimes. The simplest way to accomplish this is to use a linear interpolation algorithm to advance the field from the $ (i -1)$th slice to the $i$th slice. Using this procedure ONEDFEL  can treat electron beams and radiation fields with arbitrary temporal profiles and it is possible to simulate complex spectral properties.

Simulations, effectively, work in the lab frame,
The general set-up  consists of 
\begin{itemize}
\item   guide \Bf\  $ B_0$ (as strong as numerically possible); 
\item   a wiggler with wave number $k_w$   and relative amplitude  $\delta =   B_w/B_0 \ll 1$ is  as EM wave (with adiabatic turning-on);
wiggler's frequency in the beam frame  is {\it below} the cyclotron frequency associated with the guide field, $\om_w  \sim \gamma_b  \ll  \omega _{B_0}$ (but can be comparable to  the cyclotron frequency of the wiggler, $  \omega _{B_w}$);  
\item     charged beam with  ``solid''  (dead) neutralizing  background; the corresponding \Alfven wave is relativistic, $v_A \sim c$. The beam is initially propagating along  the \Bf\ (no gyration)
\item  pulse duration is  much longer than the wiggler wavelength.
\end{itemize}
By using a pure EM wave, and  not as a self-consistent \Alfven wave,  eliminates complications  related to setting the correct particle currents. In the highly magnetized regime $\sigma \gg 1$ \Alfven waves are nearly luminal  \citep[here $\sigma = B^2/(4\pi \rho c^2)$ is the magnetization parameter][]{1984ApJ...283..694K}.

\begin{figure}
  \centering
    \includegraphics[width=0.8\textwidth]{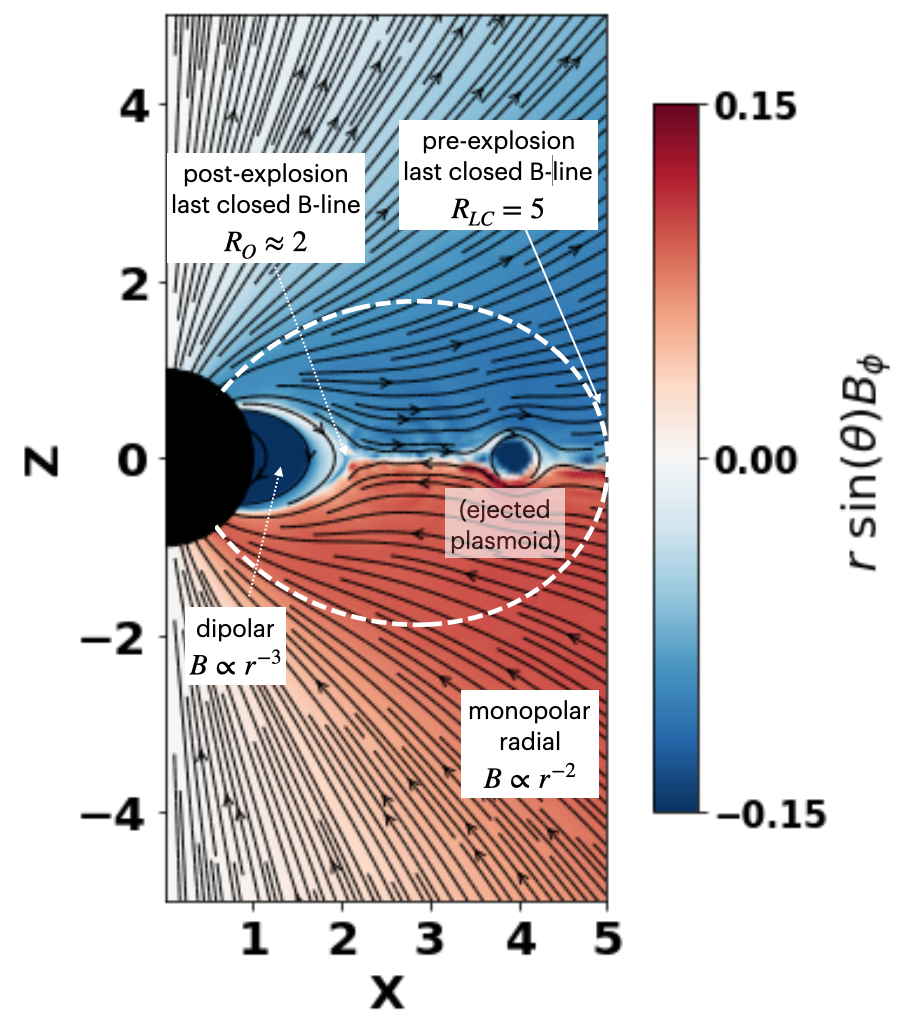}
     \caption{ Post-flare opening of the \ms\ \citep{2023MNRAS.524.6024S}. Color is values  $r \sin \theta B_{\phi}$, lines are poloidal field lines. The spin parameter is $\Omega =0.2$, so that before the ejection of a CME the \LC\ is at $R_{LC} =5$ (last closed \Bf\ line is dashed white). Post-flare \ms\ is open starting $R_O \ll  R_{LC}$. At   $ r \geq R_O $ the \ms\ has  monopolar-like \Bf\ structure.  Appearance of plasmoids ("ejected plasmoid") is not important for present work.}
\label{StreamlinexBphi_LocalShearing_Symmetric_Pi4}
\end{figure}

Let us next comment on the  applicability of the  1D regime.   The  post-eruption magnetic field lines are mostly radial. 
Development of a Coronal Mass Ejection (CME), accompanying magnetospheric FRBs,  leads to opening of the \ms\ from radii  $R_0$, much smaller than the \LC\  \citep{2023MNRAS.524.6024S}, Fig. \ref{StreamlinexBphi_LocalShearing_Symmetric_Pi4}.  After the generation of a CME the \ms\ becomes open, with nearly radial \Bf\ lines for $r\geq R_O $

\section{Results}

In simulating the magnetar magnetosphere environment we consider an electron beam propagating along the magnetic field in the presence of a plane-polarized electromagnetic wave. The basic parameters are shown in Table \ref{param}. We consider a mono-energetic 50 MeV beam over a bunch length/charge of 1.8 $\mu$s/5.9 mC with a peak current of 5 kA. The beam plasma frequency corresponding to a 5 kA beam with a radius of 100 cm is about 1.6 kHz. The electromagnetic undulator is taken to have a period of 100 m and an amplitude of 0.01 kG. The excited radiation, therefore is also plane-polarized. We study the interaction for various values of the axial field so that the resonant wavelength will vary with the axial field.

\begin{table}
\begin{center}
\begin{tabular}{|c|c|}
Beam Energy & 	50 MeV\\
Peak Current	&5000 A\\
Bunch Duration	 & 1.8 $\mu$s\\
Beam Radius	& 100 cm \\ 
Pitch Angle Spread	& 0\\ 
Period	&100 m\\
Amplitude	 & 0.01 kG \\
Polarization	& Planar\\
Axial Magnetic Field ($B_0$)	& Variable \\
Amplitude	 & Variable
\\ 
\end{tabular}
\end{center}
\caption{Parameters of simulations}.
\label{param}
\end{table}

The parameters of  simulations  nearly match the real physical condition (except the value of the guide field): for a beam Lorentz factor $\gamma_b \sim 100$  and a wiggler length $\lambda_w \sim 100$  meters the resonant wavelength is a few centimeters. These values are close to the real scales we expect in neutron star magnetospheres. As mentioned previously, the guide field is below that  expected but the numerical simulation becomes more and more computationally challenging as the resonant linewidth becomes narrower for high guide fields. However, the wavelength becomes independent of the guide field (Fig.  \ref{003}). In this particular example, over a few kilometers (also a realistic physical value) the intensity grows by fourteen orders of magnitude and reaches saturation.

\subsection{Steady-state runs}

\begin{figure} 
\includegraphics[width=.49\linewidth]{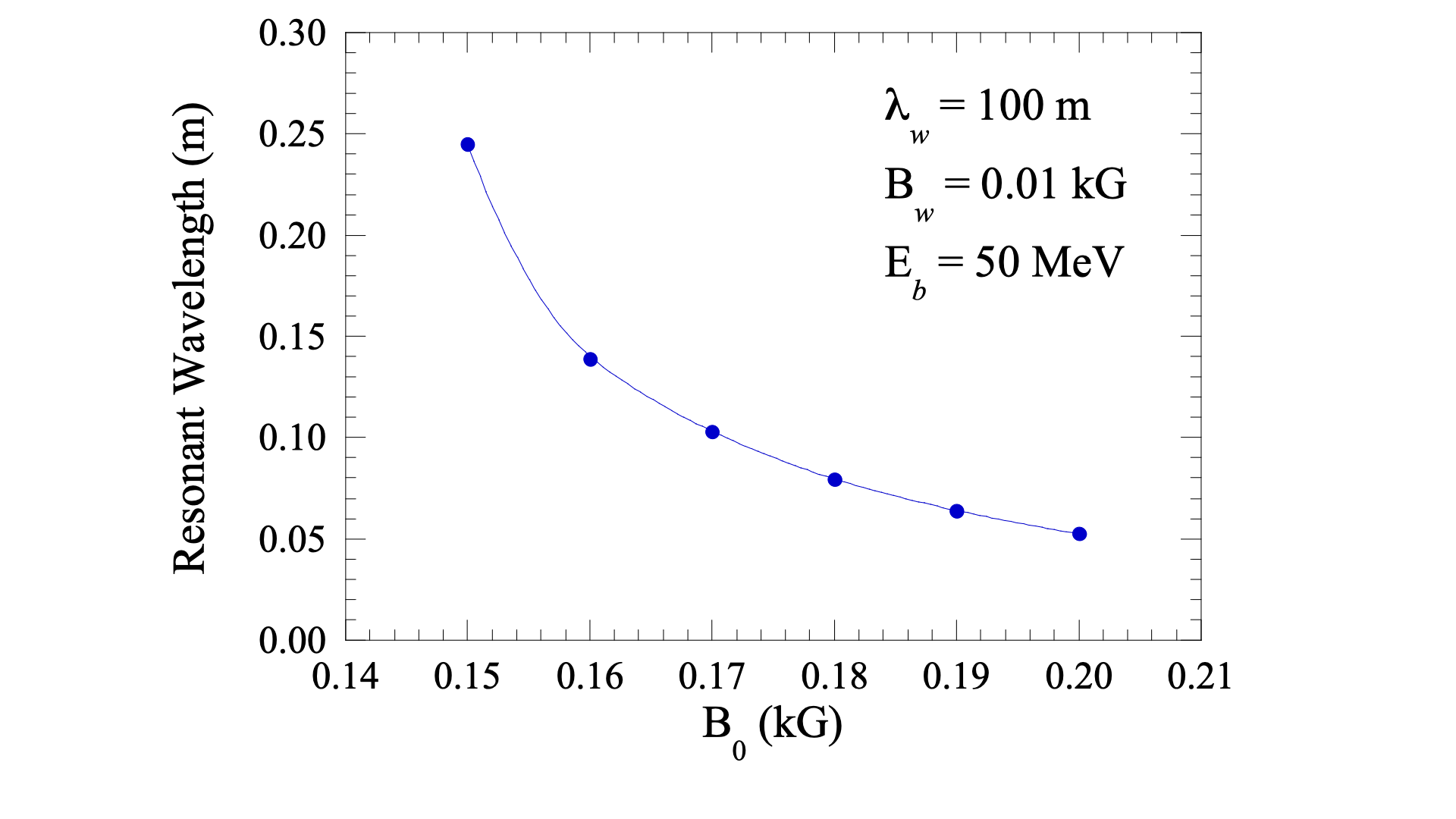}
\includegraphics[width=.49\linewidth]{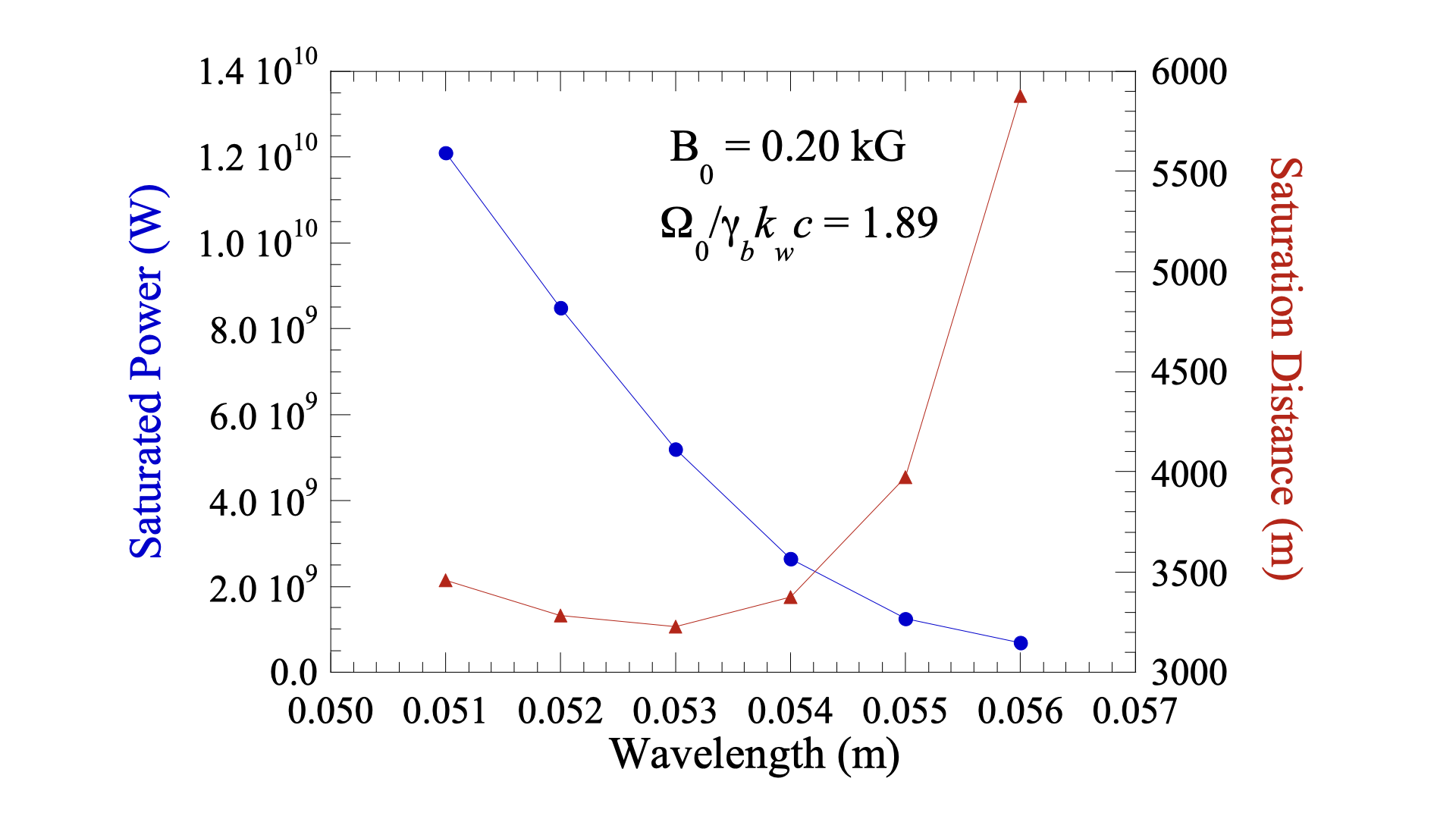}
\caption{Steady-state runs. Left panel: Variation in the FEL resonant wavelength with magnetic field in the Group II regime. Right panel: The saturated power and saturation distance versus wavelength for $B_0 = 0.2$ kG.}
\label{003} 
\end{figure}

Using steady-state (i.e., time independent) ONEDFEL simulations, we have studied the variation in the resonant wavelength with increases in the magnetic field. As shown in Fig. \ref{003}, the resonant wavelength for the FEL interaction decreases from about 0.25 m for a magnetic field of 0.15 kG to 0.053 m when the magnetic field increases to 0.20 kG. We observe that the curve is approaching an asymptote as $B_0$ increases past 0.20 kG. This means that the resonant wavelength will remain relatively constant as the magnetic field increases above this value and we expect that the interaction properties will not change significantly for still higher field levels. This is important because simulations become increasing challenging as the field increases beyond this point. {\bf  Independence of the resonant wavelength on the value of the guide \Bf\ is expected, see  Eq. (\ref{omC}). }

The variation in the saturated power and saturation distance (when starting from noise) are shown in Fig. \ref{003}  for $ B_0 = 0.20$ kG. Here we observe that the full width of the gain band extends from about 0.051 m to 0.056 m and the optimal wavelength, corresponding to the shortest 
saturation distance is 0.053 m (as indicated in Fig. \ref{003}) and that the decreases rapidly as the wavelength increases within this gain band.

\subsection{Time-dependent simulations}

 \begin{figure}
\includegraphics[width=.99\linewidth]{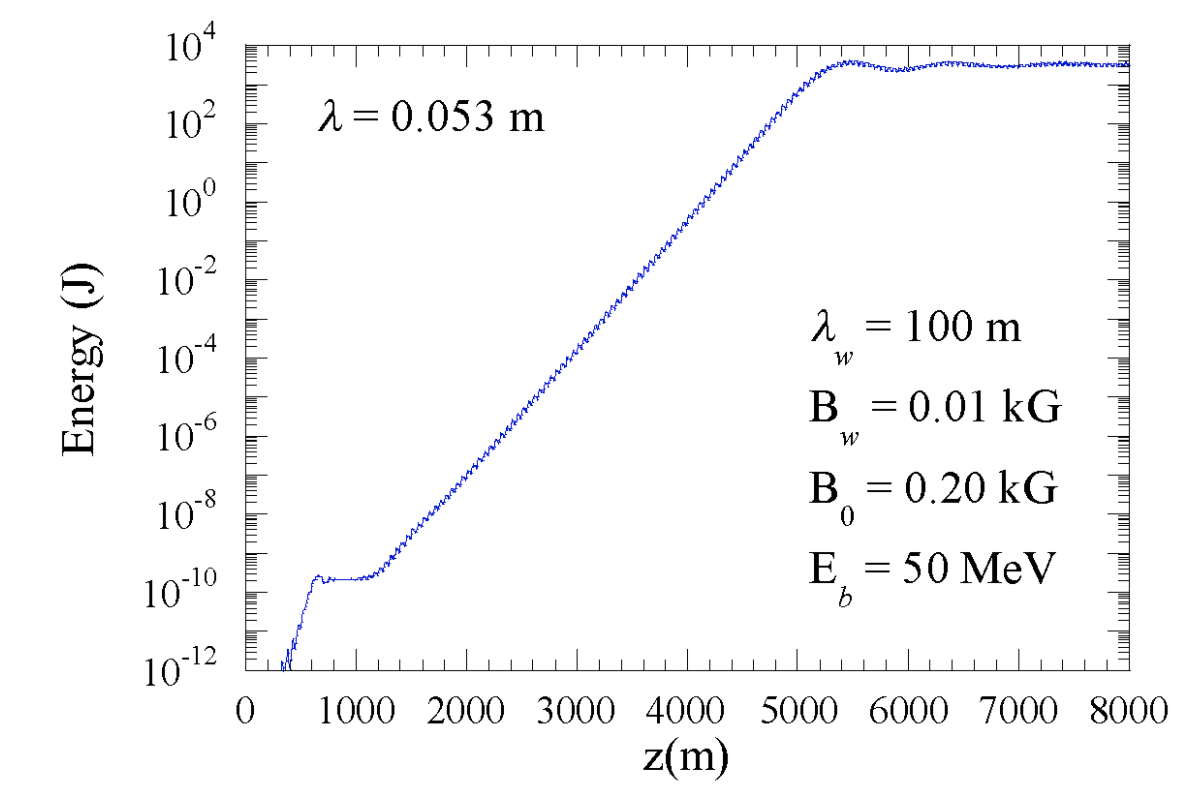}
\caption{ 
FEL power as a function of distance.  {\it We observe a growth  of the EM signal  over $\sim$ fourteen orders of magnitude, from noise (no seeding)!}  (over a scale of $\sim$ few kilometers - realistic \NS\ setting).}
\label{ONEDFEL1} 
\end{figure}

Next, we ran  time-dependent simulations using ONEDFEL. Simulations were conducted to determine the resonant wavelengths for different values of the axial magnetic field. Note that while these are 1D simulations, we need to specify the beam radius in order to determine the current density and beam plasma frequency.

Our  results are plotted in Fig. \ref{ONEDFEL1}. Importantly, the signal evolves from noise -  there is no seeding. The simulations  nearly match the real physical condition:  for beam \Lf\ $\gamma \sim  100$ and wiggler length $\lambda_w =100 $ meters 
the  resonant wave length is  few centimeters!   These  values are close to the real scales we expect in \NS\ \mss!).  The guide field is below that of  the expected, though.   For high guide fields the procedure  becomes more and more computationally challenging as the resonant line becomes narrower.   Over a scale of few kilometers (also a realistic physical value) the intensity grows by fourteen orders of magnitude, reaching saturation.

\begin{figure}
\includegraphics[width=.99\linewidth]{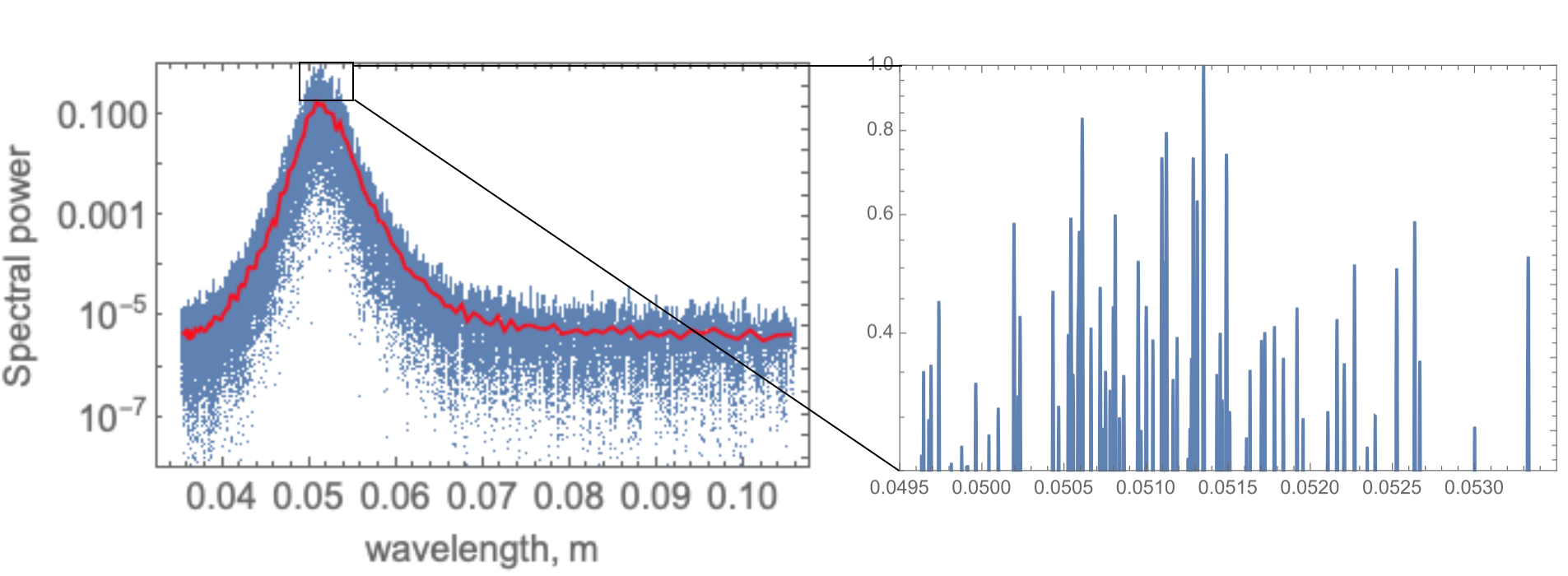}\\
\includegraphics[width=.99\linewidth]{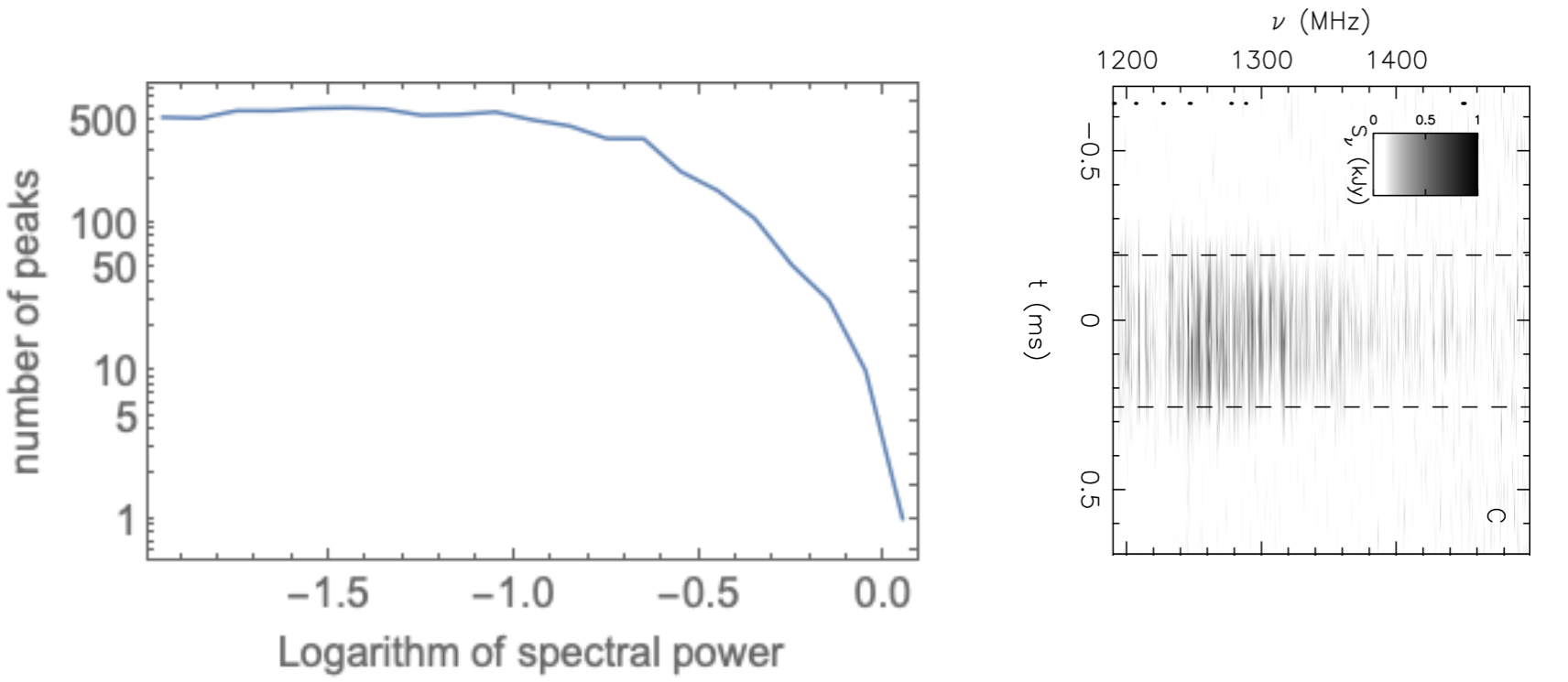}
\caption{ Top row:   Time-dependent simulations of the FEL with ONEDFEL  showing  a line-like feature (left)  with zoomed-in  (right) complicated  internal spectral structure of FEL in SASE regime.  Red line is averaged spectrum.  Bottom row, left panel: distribution of brightest  peaks normalized to maximal; Bottow right:  Dynamic spectrum of FRB150807 \protect\citep{2016Sci...354.1249R} showing  strong modulations in both frequency and time. (Time and frequency structures are inter-dependent in the de-dispersion procedures.)}
\label{bands} 
\end{figure}

 The resulting spectral structure is most revealing, Fig. \ref{bands}. We find, first, that there is a typical wavelength of the produced emission  (top left  panel) -  this is a natuaral consequence of our assumptsion.
We  also find that the pulse has a complicated internal spectral structure,  (top right panel in Fig. \ref{bands}) - which is a natural property of a SASE FEL. The spectral width of the central spike is less than 1\% (FWHM).
This complicated internal structure  resembles what is  indeed seen in Fast Radio Bursts,  Fig. \ref{bands}, bottom right  panel.
 FRBs display a wide variety of complex time-frequency structures \citep{2016Sci...354.1249R,2018Natur.553..182M},  including strong modulations in both frequency and time (with characteristic bandwidth of $\sim 100$ kHz.)
 

We also attempted a statistical description of the peaks, bottom left panel in Fig.  \ref{bands}. Since we are not aware of any theoretical prediction for the distribution of the peak, we cannot do a proper statistical analysis. Plus, naturally, any particular realization of the peaks is subject to numerous uncertainties, both physical, numerical, and statistical.

\subsection{Connection to CARM regime}
\label{CARM}

\begin{figure} 
\includegraphics[width=.99\linewidth]{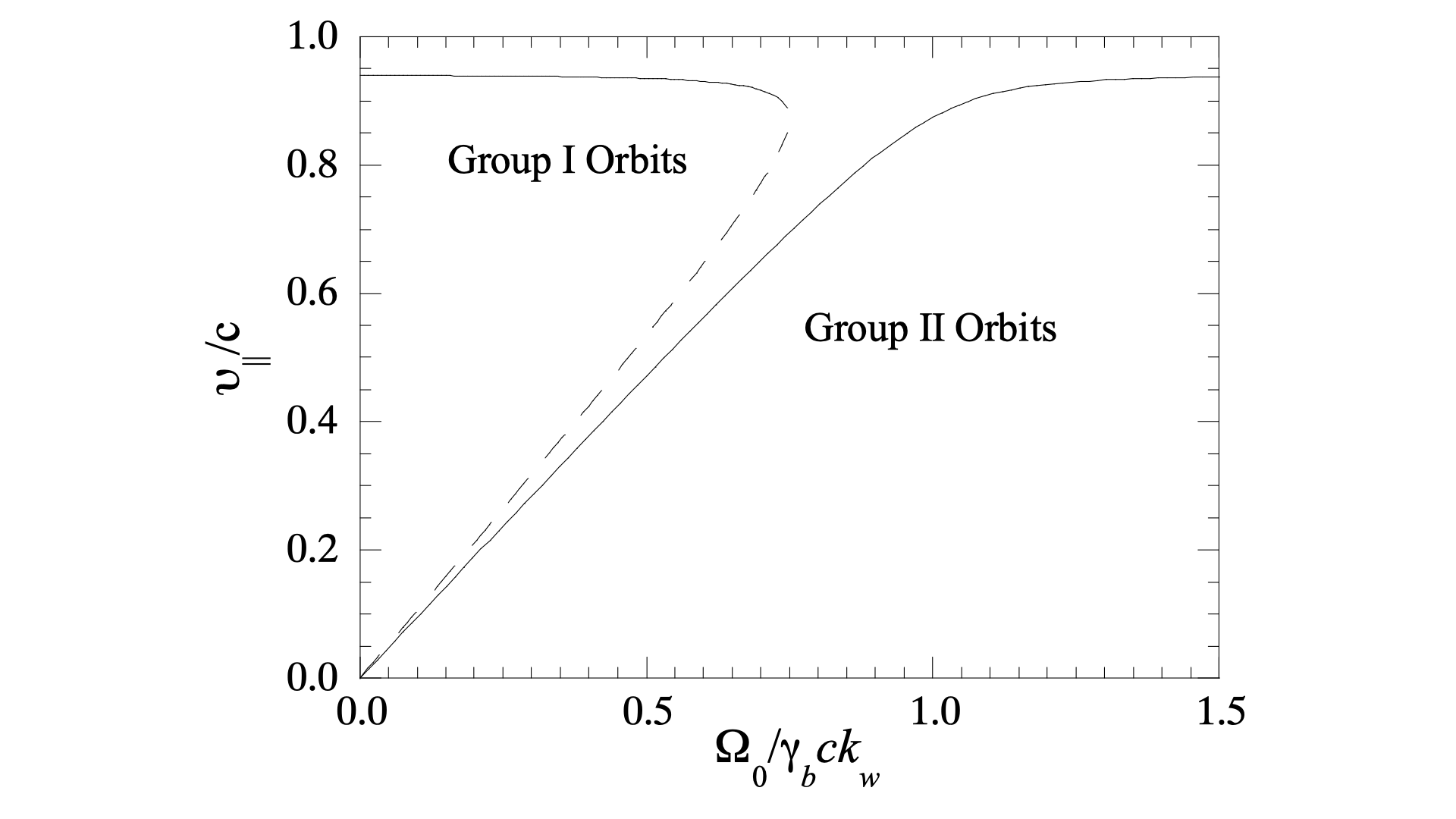}
\caption{  Illustration of the variation in the axial velocity with the axial magnetic field for Group I and Group II trajectories  \protect\citep{freund1986principles}.}
\label{001} 
\end{figure}

The particle-wave interaction may lead to the excitation of the cyclotron motion and ensuing azimuthal bunching of emitting electrons. This will take the us into the Cyclotron Auto Resonance Maser (CARM regime). In this case, the resonant wavelength is governed by the axial velocity of the electron beam and, for fixed beam energy and undulator parameters, this will vary with the axial field. The combination of the wiggler and axial magnetic fields results in particle trajectories that exhibit a magneto-resonance in which the transverse velocity increases as the difference between the wiggler and Larmor periods decreases. The energy corresponding to the transverse velocity cannot exceed the total energy; hence, there are two distinct classes of trajectories corresponding to cases where the Larmor period is longer than the wiggler period (termed Group I) and shorter than the wiggler period (termed Group II). The variation in the axial velocity, $\beta _\parallel$, as the axial magnetic field increases (for given values if the wiggler period and field strength and electron energy) is illustrated in  Fig.  \ref{001}  \citep[taken from ][]{freund1986principles} where the dashed line indicates unstable trajectories. The figure is meant to show the generic variation in the average axial velocity versus the axial field strength for a magneto-static wiggler, hence, it is not meant to correspond to the parameters used in the simulation. Note that the average transverse velocity $\beta_\perp^2 =1-1/\gamma_b^2 - \beta_\parallel^2$. Group I trajectories are generally found in the weak axial field regime below the magneto-resonance ,$\Omega_0 \leq \gamma_b k_w c$, where the Larmor period is longer than the wiggler period. Group II trajectories occur in the strong axial field regime where  $\Omega_0 \gg \gamma_b k_w c$. We are most concerned with Group II trajectories in which  $\Omega_0 \gg \gamma_b k_w c$, which we expect to be relevant to the conditions in magnetar magnetospheres. As shown in Fig. \ref{001},  the axial velocity increases but asymptotes with increasing magnetic field in the Group II regime which implies that the FEL resonant wavelength decreases with increasing magnetic field but reaches a value which is relatively independent of the axial field strength.

\begin{figure} 
\includegraphics[width=.99\linewidth]{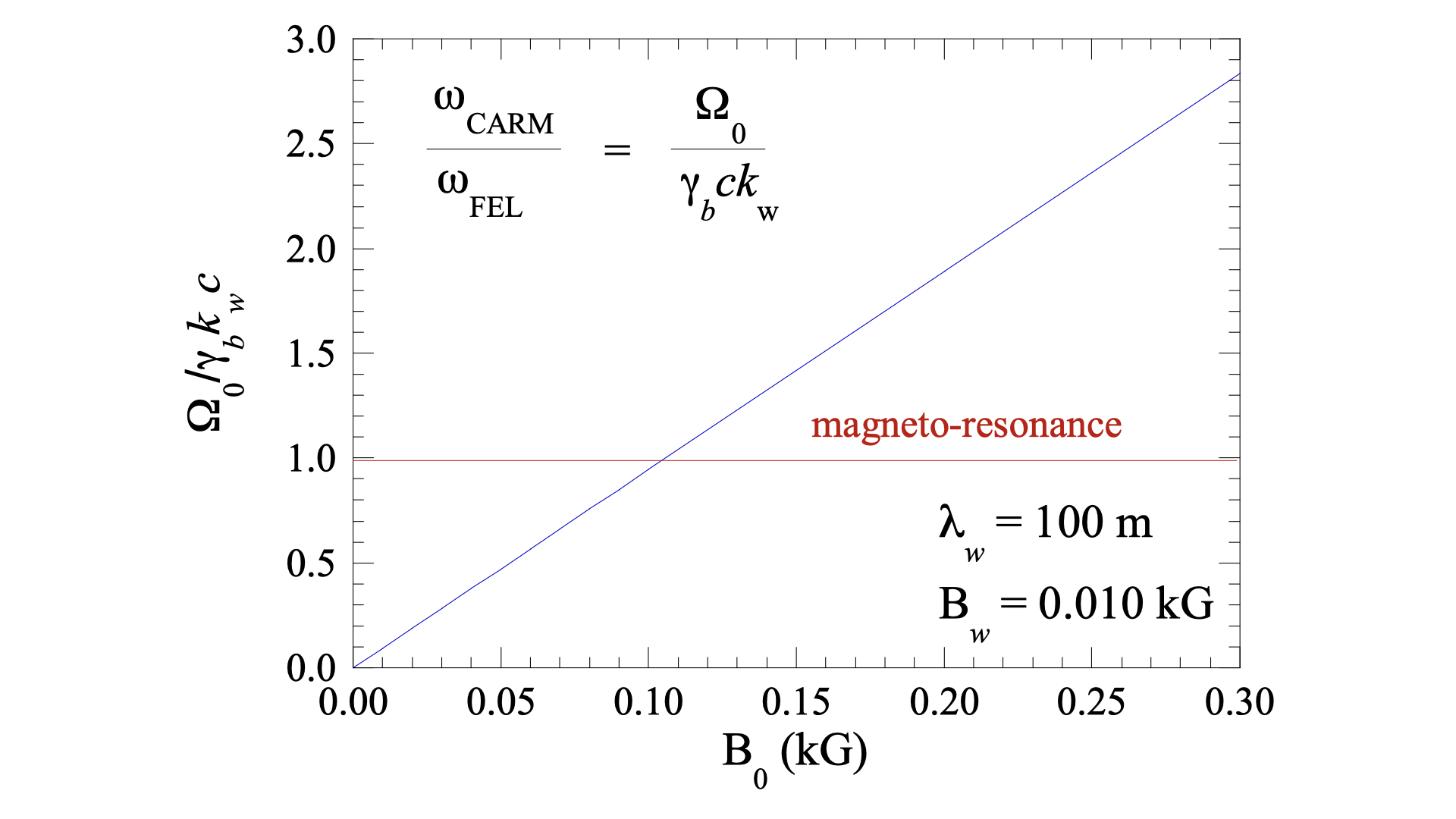}
\caption{Plot showing the variation in $\Omega_0/\gamma_b k_w c$  versus the axial field showing that Group I orbits are found when $B_0 \le 0.10$  kG and Group II orbits when $B_0 \ge $  0.10 kG. }
\label{002} 
\end{figure}

We remark that there are two possible interactions of an electron beam streaming along the field lines corresponding to the FEL resonance and that of a cyclotron auto-resonance maser (CARM). The ratio of the resonant frequencies of these two interaction mechanisms is given by
\be
\frac{\om_{CARM}}{\om_{FEL}} = \frac{\Omega_0 }{\gamma_b k_w c},
\ee
so that the FEL resonance is found at a lower frequency than that for the CARM for strong axial guide fields in the Group II regime. For the parameters of interest here, the magneto-resonance is found for an axial field of about 0.10 kG as shown in Fig. \ref{002}. We are primarily concerned here with the strong axial field regime.
This is a separate, and possibly astrophysically  important  emission mechanism. We leave the analysis of CARM regime to a separate future investigation.

\section{Conclusion}

In this work we numerically study  operation of Free Electron Laser in the astrophysically important guide-field dominated regime.  In this regime particles mostly slide along the dominant guiding \Bf\  and experience ${\bf E} \times {\bf B}$ drift in the field of the wiggler. 
 Our parameters (energy of the beam, wavelength of the wiggler) closely match what is expected in the \mss\ of \NSs. The value of the guide field is, though, much smaller than expected. However, we verified that  the wavelength becomes independent of the guide field, Fig.  \ref{003}. 

Our results are encouraging. First,  we see  {\it unseeded} growth over 14 orders of magnitude over the  real physical  scale of $\sim$ few kilometers. It is expected that the real \mss\ are much ``noisier``, with mild level of intrinsically present turbulence.  The presence of such turbulent \Alfven waves will provide seeds to jump-start the operation of the FEL. 

The most intriguing  result is, perhaps,  the fine spectral structure, Fig. \ref{bands}, that qualitatively  matches observations. Such fine structure is an inherent property of SASE FEL, as different narrow modes are amplified parametrically.

Limitations of our approach include:
\begin{itemize} 
\item One-dimensional approximation. In this case we neglect curvatures of the \Bf\ lines, and corresponding particles' trajectories. We plan to address this in a separate work, using MINERVA code. 
\item
The saturation level will be affected by the higher guiding field.
 For a single  quasi-monochromatic  wave the ponderomotive potential (\ref{H})  increases linearly with EM wave intensity $E_w$, while energy density of EM waves increases quadratically $\propto E_w^2$. The balance is achieved at 
\be
\frac{ E_{EM}}{E_{w}} = \frac{ \om_{p,b}^2}{ k_{w} c \Omega_0},
\label{sat}
\ee where $\om_{p,b}$ is beam plasma density.
This is an estimate of the saturation level of the EM waves in the beam frame. 
\item
We have not addressed the energy spread in the beam. It is expected that in  the guide-field dominated regime the operation of FEL is much more tolerant to the beam spread \cite{2021ApJ...922..166L}  since in this regime the particle trajectory is {\it independent of energy}. 
 In the broad-band case the saturation will be determined  $\sim$ by a (random phase) quasilinear diffusion. In this regime the growth rate of the EM energy of the wave due to the development of the parametric  instability, Eq. (\ref{Gamma}), will be balanced by the particle diffusion (random phases!) in the turbulent EM field (the diffusion coefficient $\propto E_w^2$). 
 \item Coherence of the wiggler. We assumed purely monochromatic wiggler. Spectral spread of the wiggler will tend to reduce the FEL efficiency.  
 \end{itemize} 
We plan to address theses issues in a  future publication.
  

This work had been supported by NASA grants 80NSSC17K0757 and  NSF grants 1903332 and 1908590. 

\bibliographystyle{apsrev}
 \bibliography{/Users/maxim/Dropbox/Research/BibTex}

  \end{document}